%Paper: hep-th/9305066
%From: Jeff Harvey <harvey@poincare.uchicago.edu>
%Date: Fri, 14 May 93 12:03:35 -0500
%Date (revised): Fri, 14 May 93 13:27:07 -0500

%-------------------------
% This paper uses harvmac
%-------------------------

\input harvmac
\noblackbox
%
% Macros to remove Harvard titlepage from harvmac
%                   S.B.G. 3/91
%

%
%---------------------------------------------------------------------
%

\def\Title#1#2{\ifx\answ\bigans \nopagenumbers
\abstractfont\hsize=\hstitle\rightline{#1}%
\vskip .5in\centerline{\titlefont #2}\abstractfont\vskip .5in\pageno=0
\else \rightline{#1}
%\abstractfont %\hsize=\hstitle
%\rightline{#1}%
\vskip .8in\centerline{\titlefont #2}%\abstractfont
\vskip .5in\pageno=1\fi}
\ifx\answ\bigans

\else

 \font\absi=cmmi10 scaled\magstep1
\font\absis=cmmi7 scaled\magstep1 \font\absiss=cmmi5 scaled\magstep1
\font\abssy=cmsy10 scaled\magstep1 \font\abssys=cmsy7 scaled\magstep1
\font\abssyss=cmsy5 scaled\magstep1 
\skewchar\absi='177 \skewchar\absis='177 \skewchar\absiss='177
\skewchar\abssy='60 \skewchar\abssys='60 \skewchar\abssyss='60
\fi
%
%-------------------
% useful definitions
%-------------------
%

\def\hf{{1\over2}}
\def\qt{{1\over4}}
\def\hs{{\hat s}}
\def\tht{{\hat t}}
\def\hu{{\hat u}}
\def\hM{{\hat M}}

\def\ajou#1&#2(#3){\ \sl#1\bf#2\rm(19#3)}

\def\frac#1#2{{#1 \over #2}}

%
%-------------------
%references
%-------------------
%
\lref\hetsol{A. Strominger, Nucl. Phys. {\bf B343} (1990) 167;
E: Nucl. Phys. {\bf B353} (1991) 565.}
\lref\MOOL{C. Montonen and D. Olive, Phys. Lett. {\bf 72B} (1977) 117. }
\lref\osborn{H. Osborn, Phys. Lett. {\bf 83B} (1979) 321.}
\lref\khurione{R. R. Khuri, {\sl A Multimonopole Solution in String Theory},
Texas A\&M preprint CTP/TAMU-33/92 (hep-th/9205051) (April 1992);
{\sl A Heterotic Multimonopole Solution}, Texas A\&M preprint
CTP/TAMU-35/92 (hep-th/9205081) (April 1992).}
\lref\BEDR{E. A. Bergshoeff and M. de Roo, Nucl. Phys. {\bf B328} (1989) 439.}
\lref\world{C. G. Callan, J. A. Harvey and A. Strominger, Nucl. Phys.
{\bf B359} (1991) 611.}
\lref\raj{R. Rajaraman, {\it Solitons and Instantons}, North-Holland,
Amsterdam (1982).}
\lref\gpy{D. J. Gross, R. D. Pisarski and L. G. Yaffe, Rev. Mod. Phys.
{\bf 53} (1981) 43.}
\lref\khuria{
R.~R.~Khuri, ``Classical Dynamics of Macroscopic Strings",
Texas A\&M preprint CTP/TAMU-80/92 (to appear in Nucl. Phys. B).}
\lref\dh{
A.~Dabholkar and J.~A.~Harvey, Phys. Rev. Lett. {\bf 63} (1989)
478.}
\lref\dghr{
A.~Dabholkar, G.~Gibbons, J.~A.~Harvey and F.~Ruiz Ruiz,
Nucl. Phys. {\bf B340} (1990) 33.}
\lref\ck{
C.~G.~Callan and R.~R.~Khuri, Phys. Lett. {\bf B261} (1991) 363.}
\lref\khurib{
R.~R.~Khuri, ``Geodesic Scattering of Solitonic Strings",
Texas A\&M preprint CTP/TAMU-79/92.}
\lref\joe{
J.~Polchinski, Phys. Lett. {\bf B209} (1988) 252.}
\lref\gsw{
M.~B.~Green, J.~H.~Schwarz and E.~Witten,
{\it Superstring Theory} vol. 1, Cambridge University Press (1987).}
\lref\heterosis{
D.~J.~Gross,
J.~A.~Harvey, E.~J.~Martinec and R.~Rohm, Nucl. Phys. {\bf B256} (1985) 253.}
\lref\fivemod{Ref on fivebrane moduli space}
\lref\iz{C. Itzykson and J-B. Zuber, {\it Quantum Field Theory}, McGraw-Hill
(1980).}
\lref\rube{ P. Ruback,  Commun. Math. Phys. {\bf 107} (1986) 93.}
\lref\landl{See L. Landau and E. Lifshitz,
{\it Classical Theory of Fields}, Fourth Revised
Edition, Pergamon Press (1975),  for a
derivation and the gravitational generalization.}
\lref\manone{N. S. Manton, Phys. Lett. {\bf 110B} (1982) 54.}
\lref\mantwo{N. S. Manton, Phys. Lett. {\bf 154B} (1985) 397; E: Phys. Lett.
{\bf 157B} (1985) 475.}
\lref\ah{M. Atiyah and N. Hitchin, {\it The Geometry and Dynamics of Magnetic
Monopoles}, Princeton University Press (1988) and references therein.}
\lref\gm{G. W. Gibbons and N. S. Manton, Nucl. Phys. {\bf B274} (1986) 183.}
\lref\khuris{R. Khuri, preprints hep-th/9212026, hep-th/9212029,
hep-th/9303074.}
\lref\flatfive{A. G. Felce and T. M. Samols, ``Low-Energy Dynamics of String
Solitons, hep-th/9211118.}
\lref\het{D. Gross, J. A. Harvey, E. Martinec, and R. Rohm,  Nucl Phys. {\bf
B267}
(1986) 75.}
\lref\pisson{In preparation.}
\lref\tye{H. Kawai, D. C. Lewellen, and S. H. Tye, Nucl. Phys. {\bf B269}
(1986) 1.}
\lref\dan{D. Waldram, Phys. Rev. {\bf D47} (1993) 2528.}
\lref\duff{M. J. Duff and J. X. Lu, Phys. Rev. Lett. {\bf 66} (1991) 1402.}
\lref\sena{A. Sen, Nucl. Phys. {\bf B388} (1992) 457.}
\lref\lmech{See e.g. L. Landau and E. Lifshtiz, {\it  Mechanics}, Third
Edition,
Pergamon Press.}
\lref\twave{J. Gauntlett, J. Harvey, and D. Waldram, in preparation}
\lref\busc{T. Buscher, Phys. Lett. {\bf B194} (1987) 59.}
%
%-------------------
% title page
%-------------------
%
\Title{\vbox{\baselineskip12pt
\hbox{EFI-93-29}
\hbox{hep-th/9305066}}}
{Scattering of Macroscopic Heterotic Strings}
{
\baselineskip=12pt
\bigskip
\centerline{Jerome P. Gauntlett,  Jeffrey A. Harvey}
\centerline{Merle M. Robinson, and Daniel Waldram}
\bigskip
\centerline{\sl Enrico Fermi Institute, University of Chicago}
\centerline{\sl 5640 Ellis Avenue, Chicago, IL 60637 }
\centerline{\it (jerome,harvey,merle,waldram)@yukawa.uchicago.edu}
\bigskip

\centerline{\bf Abstract}
We show that macroscopic heterotic strings, formulated as strings
which wind around a compact direction  of finite but macroscopic
extent, exhibit non-trivial scattering at low energies.
This occurs at order velocity
squared and may thus be described as geodesic motion on a moduli
space with a non-trivial metric which
we construct. Our result is in agreement with
a direct calculation of the string scattering amplitude.
}

%\draftmode

\Date{5/93}

%
%---------------------------------------------
%

Non-relativistic scattering or bound-state problems involving
two massive particles are conventionally studied by solving
the Schr\"odinger equation with the relevant two-body potential.
The inclusion of relativistic corrections in a systematic way requires
the use of quantum field theory, where, in the words of \iz\
 ``Accurate predictions require some artistic gifts from
the practitioner''.
Thus phenomena which would be difficult to study directly in quantum
field theory may be analyzed more simply by use of
the Schr\"odinger
equation, particularly when tunneling is
involved. In string theory we know how to calculate scattering
amplitudes for particular modes of the string, but these
typically  are either massless or massive with lifetimes
of order the Planck time and so there is no limit in which
their interaction can be described by the Schr\"odinger
equation.  A simple method of obtaining massive stable
states in string theory is to compactify the theory on a
non-simply connected space and to look at string states
with non-zero winding number.  The interaction of such states
at low velocities should be amenable to a non-relativistic treatment
which potentially contains information not directly visible
in string perturbation theory. In this paper we will make
a first step in this direction by constructing the non-relativistic
Lagrangian which describes the interaction of string
states wound around a circle.

In certain special systems, often related to supersymmetry,
the static force between two massive particles or solitons
vanishes due to cancellation of forces generated by exchange of particles
of different spin. This is known to be the case for the
string winding states we will consider \refs{\dh, \dghr}.
In this situation it is still possible to
study non-relativistic scattering and bound state problems,
but now there is no static potential and the Laplacian
appearing in the Schr\"odinger equation should be
constructed using the metric on the two particle ``moduli space''.

To illustrate this consider a complex massive scalar
$\Phi$ coupled to a massless scalar $\phi$ and a massless
vector field $A_\mu$ with Lagrangian
\eqn\one{{\cal L}_{S+V} = - \qt F_{\mu \nu} F^{\mu \nu} + (D_\mu \Phi)^*
      (D^\mu \Phi) - m^2 \Phi^* \Phi + \hf \partial_\mu \phi \partial^\mu \phi
     - 2 m f \phi \Phi^* \Phi }
with $D_\mu = \partial_\mu - i e A_\mu$. The tree-level t-channel
contribution to the invariant matrix element for $\Phi \Phi \rightarrow
\Phi \Phi$ scattering is then
\eqn\two{i {\cal M}_t = {i e^2 \over t }(2 s -4m^2 +t) - {4 i m^2 f^2 \over t}}
with $s,t,u$ the usual Mandelstam invariants. Near threshold
$s \sim 4 m^2$, $t \sim 0$ and we see that the static force
vanishes if $e^2 = f^2$. Assuming this to be the case,  we  find
\eqn\three{i {\cal M}_t = ie^2 { (2s-8 m^2) \over t} +  i e^2  .}

If we wish to use \three\ to calculate the
non-relativistic differential cross-section
for scattering at large impact parameter (and hence at small
angles)  then
we can drop the last term (which gives rise to a contact interaction)
and take the non-relativistic limit to find
\eqn\four{{d \sigma \over d \Omega} = {|{\cal M}|^2 \over 64 \pi^2 s}
	\rightarrow {e^4 \over 64 \pi^2 m^2 \sin^4 (\theta/2)}}
where $\theta$ is the scattering angle in the center of mass frame.

If we generalize \one\ to include coupling to gravity as well with
\eqn\five{S_{S+V+G} = \int d^4 x \sqrt{-g} \left( {R \over 16 \pi G} +
{\cal L}_{S+V} \right) }
then including graviton exchange we find for the t-channel contribution
\eqn\six{\eqalign{ i {\cal M}_t = {i \over t} \bigl[ & -f^2 (4 m^2) +
                                       e^2 (2s -4m^2 +t) \cr
                                    & -16 \pi G (m^4 - 2 m^2 s -m^2 t + s^2/2
					+st/2 )
                                     \bigr] . \cr }  }
The static force now cancels if
\eqn\seven{-f^2 + e^2 - 4 \pi G m^2 =0 , }
and if
\eqn\eight{e^2 = 16 \pi G m^2}
then the $O(v^2)$ force (and the contact terms)
will also vanish. In this case exchange of
massless fields leads to a flat metric on moduli space.
This happens e.g. for the Kaluza-Klein monopole \rube\ and for the
neutral fivebrane \refs{\ck, \flatfive}.

To  describe theories in which the static force vanishes in the
language of non-relativistic quantum mechanics we should first
derive the Lagrangian which governs the interaction of
slow-moving particles and then quantize the resulting theory.
Since the static potential vanishes, it is crucial to include interactions
up to quadratic order in velocities. This may be done by
calculating the appropriate retarded potential seen by one particle
due to the motion of the other. Assuming the particles
have equal electric charge $e$ and mass $m$ this  leads to \mantwo
\eqn\nine{ {\cal L} = \hf m v_1^2 + \hf m v_2^2 - {e^2 \over 8 \pi r}
                 ( \vec v_2 - \vec v_1)^2 }
where $\vec r$ is the separation of the two particles.
The Lagrangian \nine\  is a generalization of the
Darwin Lagrangian of electromagnetism \landl.
Note that back-reaction effects due to radiation are higher
order in $v/c$ and hence may be dropped to this order.

It is useful to think of the Lagrangian \nine\ as describing motion
on the ``moduli space'' $R^3 \otimes (R^3-\{0\})$ of two-particle
configurations with the metric for the center of mass motion being
flat, and the metric for the relative motion given by
\eqn\ten{g_{ij} =  \left({m \over 4} - {e^2 \over 8 \pi r} \right) \delta_{ij}
{}.
		 }
The quantum problem involves study of the Schr\"odinger equation
constructed using the Laplacian for the metric \ten\ and may be used to study
phenomena beyond the simple Born approximation scattering given by \four.

This language is more commonly used to describe soliton
scattering \manone. In particular,
the scattering of BPS monopoles has been treated in some detail
\refs{\ah, \gm}. The two-monopole moduli space has the form
\eqn\eleven{{\cal M} = R^3 \times \left ( {S^1 \times M_{AH} \over Z_2}
	\right ) }
where $R^3 \times S^1$ labels the center of mass and an angle
which specifies the total electric charge. ${\cal M}_{AH}$ is
a four-dimensional manifold whose points label the relative
separation of the monopoles and a relative phase. ${\cal M}_{AH}$
has a hyperk\"ahler and hence self-dual metric which has
been determined explicitly by Atiyah and Hitchin \ah. For scattering
at large impact parameter the Atiyah-Hitchin metric reduces to
a singular form of the Euclidean Taub-Nut metric. The Taub-Nut approximation
is precisely what is found by generalizing the Lagrangian \nine\ to the
scattering of both magnetically and electrically charged dyons
in the BPS limit \mantwo, and in fact the differential cross-section
\four\ agrees with that found for pure monopole scattering in
the Taub-Nut approximation in \gm.

Although particle and soliton scattering can both be described this
way, there are some important differences. Solitons occur as non-singular
and non-perturbative
solutions to the classical field equations and the soliton moduli
space metric is complete  and non-singular. Although in practice
it has been determined indirectly, in principle the soliton moduli space
metric is determined by the kinetic part of the Lagrangian of the
underlying field theory.  Point particles on the other hand give rise
to singular solutions of the classical field equations with the
singularities determined by source terms in the usual way. There
is no apparent reason why the particle moduli space metric should
be non-singular.

We would now like to apply the above ideas to heterotic string theory.
It is known that macroscopic heterotic strings also have vanishing
static force between parallel strings of the same orientation \refs{\dh,
\dghr}. In conventional string theory macroscopic heterotic strings
occur as states in the first-quantized spectrum and lead to singular
solutions of the massless field equations, much as an electron does
in electrodynamics. On the other hand, there has been speculation that
fundamental
strings might appear as solitons in some ``dual'' formulation
of the theory \refs{\dh,\dghr,\hetsol,\duff,\sena}.  It might then
be expected that the moduli space metric we construct is only
an approximation to some non-singular metric arising in the
dual formulation, much as Taub-Nut is an approximation to the
Atiyah-Hitchin metric.

In light
of the previous discussion it is therefore of some interest to
determine the metric on the moduli space ${\cal M}_{HS}$ of macroscopic
heterotic string configurations. The metric on ${\cal M}_{HS}$ has
been studied in previous work \refs{\ck, \khuris} where it was found to
be flat, much like the Kaluza-Klein monopole \rube\ or the
neutral fivebrane \flatfive. There are several differences between
this work and that of \ck\ and \khuris. First of all,  in a direct calculation
of the string amplitude it is important to use the heterotic string
rather than the bosonic string, and to look at the behavior near
$t \sim 0$ rather than as $t \rightarrow -\infty$. This
leads to  a vanishing static force for identical heterotic
strings, but a non-zero
interaction at order $v^2$, while  bosonic
strings  feel a non-zero static force. Secondly, we will argue that the
metric on ${\cal M}_{HS}$ to leading order in $(2 \pi R)^{-1}$ (the inverse
of the length
of the string) is flat in agreement with \ck, but that there is a
subleading term which is
non-trivial. It is therefore important to do the calculation with $R$ finite
and to consider the limit $R \rightarrow \infty$ only at the end of the
calculation.

In order to study the metric on ${\cal M}_{HS}$ we first
examine the string $4$-point amplitude. We then derive
the string analog of the Darwin Lagrangian
(or Lorentz-Droste-Fichtenholz
Lagrangian in the case of gravity) which governs the motion of
two parallel macroscopic strings to order $v^2$ and from this
extract the asymptotic form of the metric on ${\cal M}_{HS}$.

To describe macroscopic string states in ten spacetime
dimensions we follow the usual procedure of taking one of
the spatial coordinates, say $X^9$, to be a coordinate on
a circle of radius $R$ and to consider string states with
non-zero winding about this circle. For the heterotic string the
ground state in the winding number one sector also has
one unit of momentum along the string and is in the
right-moving ground state \dh. The right-moving ground state consists
of a spacetime vector plus spinor.  We focus on the
vector state, in which case the winding state is described by a polarization
vector $\zeta_\mu$ with $\zeta_\mu k_R^\mu =0$.

The string coordinates can be written as
\eqn\twelve{X^0 = {\hat M} \tau  , \qquad X^9 = ({\tau \over R} + 2 R \sigma) ,
                              \qquad X^i =0 ,}
where $(\tau, \sigma)$ are world-sheet coordinates and
${\hat M}^2 = 4(R-1/2R)^2$. The scattering of two such states is
described by incoming left- and right-moving momenta
\eqn\thirteen{ {p_1^\mu}_{L,R} = (\hf {{\hat p}_1}^{\hat \mu} , {1 \over 2R}
\pm R ) ,
\qquad {p_2^\mu}_{L,R} = (\hf {{\hat p}_2}^{\hat \mu} , {1 \over 2R} \pm R ), }
and outgoing momenta
\eqn\fourteen{ {p_3^\mu}_{L,R} = (\hf {{\hat p}_3}^{\hat \mu} ,
 -{ 1 \over 2R}  \mp R ) ,
\qquad {p_4^\mu}_{L,R} = (\hf {{\hat p}_4}^{\hat \mu} ,
 -{1 \over 2R} \mp R ) ,  }
where the ${\hat p}$ are nine-dimensional momenta which in the
center of mass frame take the form
\eqn\fifteen{{{\hat p}_1}^{\hat \mu}= E(1,\vec v) , \qquad
                  {{\hat p}_2}^{\hat \mu}= E(1, -\vec v), \qquad
                  {{\hat p}_3}^{\hat \mu}= - E(1, \vec w), \qquad
                  {{\hat p}_2}^{\hat \mu}= -E(1, -\vec w),  }
with $|\vec v| = |\vec w| \equiv v$, $\vec v \cdot \vec w = v^2 \cos \theta$,
and $E^2 (1-v^2) = {\hat M}^2$. It is also useful to define the
nine-dimensional Mandelstam invariants
\eqn\sixteen{\eqalign{\hs  & \equiv  -({\hat p}_1 + {\hat p}_2)^2 \sim
        4 {\hat M}^2 (1+v^2 + \cdots ) \cr
    \tht  & \equiv  -({\hat p}_2 + {\hat p}_3)^2  \sim
        -2 {\hat M}^2 v^2 (1+ \cos \theta) + \cdots \cr
    \hu  & \equiv  -({\hat p}_1 + {\hat p}_3)^2 \sim
        -2 {\hat M}^2 v^2 (1- \cos \theta) + \cdots
    \cr }}
where we have also given their asymptotic behavior for small
velocities.

It is now straightforward to calculate the $4-$point string amplitude
using the techniques described in \refs{\het, \tye} with the result
\eqn\seventeen{ A_4 \propto
    {\Gamma(3 + \hM^2/2 - \hs/8) \Gamma(-\tht/8) \Gamma(-\hu /8) \over
    \Gamma(2+\hM^2/2 -\hs/8 -\tht/8) \Gamma(1-\tht/8 -\hu/8) \Gamma(2+\tht/8)}
     K(1,2,3,4)}
where $K(1,2,3,4)$ is a kinematic factor familiar from the open
superstring,
\eqn\eighteen{\eqalign{ K = -{1 \over 64} \bigl [ &
    \tht (\hs - 4 \hM^2) (\zeta_1 \cdot \zeta_3) (\zeta_2 \cdot \zeta_4)
    + \hu (\hs -4 \hM^2) (\zeta_2 \cdot \zeta_3) (\zeta_1 \cdot \zeta_4)   \cr
    + &
    \tht \hu (\zeta_1 \cdot \zeta_2) (\zeta_3 \cdot \zeta_4)
    + \cdots  \bigr ] \cr }}
where we have not indicated terms in $K$ which vanish when the polarization
vectors are all taken to be transverse to both the string and the
plane of scattering.

As pointed out in \ck\ for the analogous amplitude in bosonic string theory,
\seventeen\ vanishes for all but forward or backward scattering in the
limit $R \rightarrow \infty$ with $v^2$ fixed due to the usual high-energy
behavior of
string amplitudes. However if we wish to study scattering at large impact
parameter and low velocities we can also consider taking $R$ large
but with $R v \sim 1$. In particular, if $|\tht| \le 1$ then the scattering
will be dominated by the $\tht=0$ and $\hu=0$ poles of \seventeen,
i.e. by massless particle exchange. In this limit we find
\eqn\nineteen{\eqalign{ A_4 \sim {1 \over \hu \tht} K \sim  &
    -{\hs - 4 \hM^2 \over \tht} (\zeta_2 \cdot \zeta_3)(\zeta_1 \cdot \zeta_4)
    -{\hs - 4 \hM^2 \over \hu} (\zeta_1 \cdot \zeta_3)(\zeta_2 \cdot \zeta_4)
    \cr
    &  + (\zeta_1 \cdot \zeta_2)(\zeta_3 \cdot \zeta_4) + \cdots  \quad .
    \cr }}
Since $(\hs - 4 \hM^2)$ is order $v^2$ we see that the static long range
force does cancel, but that there is a residual interaction of order $v^2$.
Note also that the leading terms in \nineteen\ are independent of $R$.
A similar calculation for winding states in the bosonic string shows that
they do not satisfy a no-force condition, presumably because the tachyon
modifies the solution of \dghr. The analogous calculation for macroscopic
type II superstrings shows that both the constant and the order $v^2$
terms cancel, indicating both a vanishing static potential and a flat
moduli space.
Thus the heterotic string occupies a unique position
in having no static force, but yet having a non-trivial metric on the
two string moduli
space.

We would now like to show that this result is in agreement with a direct
evaluation of the metric on moduli space for scattering at large
impact parameter. We can thus focus  on interactions mediated
by the massless modes of the string.
The massless fields outside the macroscopic string configuration
given by \twelve\ may be found by solving the equations of motion
which follow from the spacetime Lagrangian
\eqn\twenty{ S = \int d^{10} x
           e^{-2 \phi} \left( R + 4 (\partial \phi)^2 - {1 \over 3} H^2 \right)
	+ S_\sigma}
with the source term $S_\sigma$ constructed out of the string
sigma-model action
\eqn\twone{ S_\sigma = - {1 \over 2 \pi} \int d^2 \sigma
\left( \sqrt{-h} h^{mn} \partial_m X^\mu \partial_n X^\nu g_{\mu \nu}
+ 2 \epsilon^{mn} \partial_m X^\mu \partial_n X^\nu B_{\mu \nu} \right) . }

The source term is given by the configuration for a moving ground
state winding string:
\eqn\twtwo{\eqalign{ X^0 =&  E \tau \cr
    X^9 =& \tau/R + 2 R \sigma \cr
    X^i =& E y^i(\tau) \cr }}
where we have chosen a parameterization such that writing ${\dot y}^i=v^i$
we have $E^2(1-v^2)={\hM}^2$ as before.
In evaluating the source terms it is also crucial to include the relative
shift in the left- and right-moving zero point energies, that is to
use the quantum Virasoro constraints. This is equivalent to writing
for the world-sheet metric
\eqn\twthree{\eqalign{T_{mn}&\equiv
    \partial_m X^\mu \partial_n X^\nu g_{\mu \nu}-{1\over 2}
    h_{mn}(h^{pq}\partial_p X^\mu \partial_q X^\nu g_{\mu \nu})\cr
    &= J_{mn}} }
where $J_{mn}$ is a two by two matrix with elements equal to $2$ in the
case where $h_{mn}$ is conformally flat.
We expand the fields as
\eqn\twfour{\eqalign{ e^{-2 \phi} &= \overline {e^{-2 \phi}} + \Phi \cr
    B_{\mu \nu} &= \overline {b_{\mu \nu}} + b_{\mu  \nu} \cr
    g_{\mu \nu} &= \overline {g_{\mu \nu}} + h_{\mu  \nu} \cr }}
where the barred quantities are the fields of the static solution and
are given by \footnote{$\dagger$}{The solution for $R \rightarrow \infty$
was is given in \dghr, the generalization to finite $R$ was found in
\dan, and the fact that this solution correctly matches on to the
string source terms is described in \twave.}
\eqn\twfive{\eqalign{ \overline {b_{09}} &= \hf \overline  {e^{2 \phi}} \cr
    \overline {g_{\mu \nu}} &= \left( \matrix{ { \overline{e^{2 \phi}}
        \left( \matrix{ -1+D & D \cr
                           D & 1+D \cr } \right) } & \vec 0 \cr
        \vec 0 & \eta_{ij} \cr} \right)
        }}
with
\eqn\twsix{\eqalign{  \overline{e^{-2 \phi}} &= 1 + {2 \alpha  \over
    r^6 } \cr
    D &= {1 \over 2 R^2} {2 \alpha  \over r^6 } \cr }}
where $r=|x^i - X^i(x_0/E) |$, $\alpha = 1 / (12 \pi  \omega_7)$,
and $\omega_7$
is the volume of the unit seven sphere.

Although we are working in units with $c=1$, the expansion we wish
to perform can be viewed as in expansion in $1/c$ as discussed
in \landl\ and we will use this language when convenient. Specifically, the
corrections to the static fields are order $1/c^3$ and we are assuming
that  the dimensionless quantities $v^i/c \ll 1$ and $2 \alpha /(c^2 r^6) \ll
1$
(that is small velocities and large impact parameter).  It is
then straightforward,
although tedious, to solve for the corrected fields to order $1/c^4$.
We find that the non-zero components to this order which are
needed to construct the effective action to order $v^2$ are
\eqn\twseven{\eqalign{ b_{i9} &= {\alpha \over 2 r^6} \left[
    {v^i } + 6 ({ \vec v  \cdot \hat r }) \hat r_i \right] \cr
    h_{00} &= {\alpha \over r^6} \left[ -{1 \over R^2 }
        \left({v }\right)^2 + 6 \left(1 + {1 \over 2 R^2} \right)
            \left({ \vec v \cdot \hat r }\right)^2
        \right] \cr
    h_{99} &= {\alpha \over r^6} \left[{1 \over R^2}
        \left({v }\right)^2 + 6 \left(1 - {1 \over 2 R^2} \right)
            \left({ \vec v \cdot \hat r }\right)^2
        \right] \cr
    h_{i0} &= {\alpha \over r^6} \left[ \left(-1 + {3 \over 2 R^2} \right)
        \left({v_i }\right) - 6 \left(1 + {1 \over 2 R^2} \right)
            \left({\vec v \cdot \hat r }\right) \hat r_i
        \right] \cr
    h_{i9} &= {\alpha \over r^6} \left[ {3 \over 2 R^2}
        \left({v_i }\right) - 6 {1 \over 2 R^2}
            \left({\vec v \cdot \hat r }\right) \hat r_i
        \right] . \cr }}

We next construct a Lagrangian determining the motion of a second winding
string propagating in the background fields \twfour-\twseven\ produced
by the first string. Using
primed variables to represent
the second string we have for its action
\eqn\tweight{ S = -{1 \over 2 \pi}  \int d^2 \sigma^\prime
 \left( \sqrt{-h} h^{mn} \partial_m {X'}^\mu \partial_n {X'}^\nu g_{\mu \nu}
+ 2 \epsilon^{mn} \partial_m {X'}^\mu \partial_n {X'}^\nu B_{\mu \nu} \right)
.}
Since the strings only scatter in directions transverse to the winding
direction
it is useful to dimensionally reduce the action of the primed string in the
background of the first string to a nine-dimensional particle action.  In
doing this we
must be careful to choose a primed
string configuration such that the equations of
motion of the string properly include the effects of the momentum
along the string and reduce to the correct
particle equations of motion.

Demanding that the string has unit winding and no massive
oscillators excited, we can choose worldsheet coordinates
so that
\eqn\twnine{\eqalign{ {X'}^{\hat{\mu}}=& {X'}^{\hat{\mu}} (\tau ) \cr
		{ X'}^9 =& 2 R \sigma + f(\tau ) \cr
		{h}_{\tau \sigma} =& 0 \cr
		{h}_{\tau \tau} =& {h}_{\tau \tau} (\tau ) \cr
		{h}_{\sigma \sigma} =& {h}_{\sigma \sigma} (\tau ) . \cr}}
Since ${h}_{mn}$ is not conformally flat in these co-ordinates
the Virasoro constraint equations \twthree\ take a different form.
Using the fact that ${J}_{mn}$ transforms as a worldsheet tensor
we have
\eqn\thirty{ {T}_{mn}=\left[\matrix{2e^2 & 2e \cr 2e & 2 \cr}\right]}
where $e = \sqrt{[-{h}_{\tau \tau}/{h}_{\sigma \sigma}]}$.

We next substitute
\twfour-\twseven\ and \twnine\ into \tweight\ and then integrate over
$\sigma$ to obtain a particle-like action depending on ${X'}^{\hat{\mu}} (\tau
)$,
$f(\tau )$, and an ``einbein'' $e$ and coupled to the spacetime
fields $g_{\mu \nu}$ and $B_{\mu 9}$.
There are now two subtleties to take into
account to ensure that the constraints \thirty\ are properly
satisfied. First,
the mass shell condition for this ten-dimensional
particle action ( the $e$ equation of motion)
is not correct as it stands:
it must be replaced with the diagonal constraint in \thirty. Secondly,
we must enforce the off diagonal constraint in \thirty.
This constraint requires that the conserved momentum $P^9$ equals $1/R$
(that this is the correct value for the winding states
we are considering should not be surprising; it is the
result of imposing unit winding and no oscillators
in \twnine\ and the quantum Virasoro conditions in \thirty.)
Since none of the spacetime fields depend on ${X'}^9$
we can remove this constraint by  using Routh's procedure \lmech\
(essentially a consistent way of substituting the values of cyclic
coordinates  back into a Lagrangian).

Doing this we are
left with the
following action:
\eqn\thone{\eqalign{ S=-\hf   \int d\tau  & \bigl [
e^{-1} \bigl ( {\dot{X'}}^2 - { ({\dot{X'}}^{\hat{\mu}}
{g}_{\hat{\mu} 9})^2 \over  {g}_{99} } \bigr )
+ {\dot{X'}}^{\hat{\mu}}  ( {2 g_{\hat{\mu} 9} \over R {g}_{99} }
 -8R {B}_{\hat{\mu} 9} )  \cr
& - e(4{R}^{2}{g}_{99} -4 + {{1}\over{R^2}}{{1}\over{{g}_{99}}})
  \bigr ] }}
describing a massive particle in nine dimensions interacting
with a background $U(1)$ gauge field and scalar field.
After eliminating
$e$ using its equation of motion, we can expand out the action and
substitute for the fields from \twseven.  Writing ${X'}^{0}= t$
and ${\dot{X'}}^{i}={v'}^{i}$, and  choosing the gauge $t=\tau$,
we find
\eqn\thtwo{\eqalign{ S=  \int dt \bigl [
 &  \hf \hM ({{v'}^{2} }) +
{{1}\over{8}}\hM({{v'}^{2} })^{2} -
{{2}\over{R}}{{\alpha }\over{{r}^{6}}}({{v'}^{2} }) +\cr
&{{4}\over{R}}{{\alpha }\over{{r}^{6}}}({ v \cdot v'  })
\bigr ] + O(1/c^5).}}

In order to obtain the correct Lagrangian describing the motion
of the two strings to this order, we must add the terms which make
the action symmetric under the interchange of the two strings. Following
this procedure, we finally find for the
complete action to this order
\eqn\ththree{\eqalign{ S =\int d t  & \bigl [
    \hf   \hM  v^2 + \hf   \hM  {v'}^2
    + {1 \over 8}  \hM  { v^4 }
    + {1 \over 8}  \hM  {{v'}^4 }  \cr
    & - { \alpha \over r^6} {2 \over R} (\vec v - \vec v')^2
    \bigr ]  . \cr } }
Thus, after separating out the relative motion,
we see that there is a non-trivial metric on ${\cal M}_{HS}$
given by
\eqn\thfour{ g_{ij} = \left( {  \hM  \over 4}
    - { \alpha \over r^6} {2 e^{2 {\phi}_0}\over R}
\right) \delta_{ij}.    }
In this last equation, we have explicitly reinstated the
factor of  the string coupling
constant $g_s^2 \equiv e^{2 {\phi}_0}$ which was
previously taken to be unity. We then note that the metric \thfour\
is manifestly invariant under the duality symmetry  $m \leftrightarrow n$,
$R \rightarrow 1/2R$, $\phi_0 \rightarrow \phi_0 - {1 \over 2} \ln  2R^2 $
\busc\ as
expected since the first-quantized string winding state has $m=n=1$.

The metric \thfour\ should lead to scattering which in the Born
approximation agrees with the string amplitude \nineteen\ up to
the polarization dependence which we have not yet included.
We will assume that it suffices to look at scattering in which the
polarization does not change and the polarizations of the two
incoming strings are orthogonal to each other and to
the scattering plane. Then
to compare \nineteen\ to \thfour\ we take the t-channel term
in \nineteen, divide by $(2 \hat M)^2$ to change from relativistically
normalized states to the usual normalization of non-relativistic
quantum mechanics,  divide by $2 \pi R$ to obtain a nine-dimensional
matrix element, and Fourier transform to obtain a $v^2$ perturbation
to the non-relativistic Hamiltonian.  This gives
\eqn\thfourhalf{ \delta H \propto {e^{2 \phi_0} p^2 \over R {\hM}^2  r^6} }
in agreement with \thfour\ up to numerical prefactors which  were ignored
in \nineteen.

Although we have only determined the form of the metric at large $r$,
it is possible to determine the form of the corrections due to the
massive string states as well. This would clearly be an onerous
task if we were to proceed as above but with higher order in
$\alpha'$ corrections to the spacetime Lagrangian. On the
other hand the string amplitude  \seventeen\ gives the exact
contribution to the scattering amplitude from all the massive modes
of the string. Expanding \seventeen\ about the t-channel poles
at $\hat t = 8n$ it is easy to see that the no-force condition
continues to be satisfied for all the massive contributions.

What is less expected is that the
{\it first} massive pole at $\tht = 8$ has a residue of order
$v^2$, and thus contributes to the metric on moduli space,
while the residues of all the higher poles are of  order $v^4$ and
do not contribute.  At present we have no good understanding
of this rather mysterious fact.

As we mentioned earlier, in general one only expects to
find an effective Lagrangian to order $1/c^2$ because
back-reaction effects from radiation start to appear at order
$1/c^3$.  As emphasized in \landl\ there are however exceptions to this.
In purely gravitational systems radiation terms do not appear until
order $1/c^5$ and one can thus derive an effective Lagrangian to
order $v^4$. In addition, there is no electromagnetic dipole radiation for a
system of charges with equal charge to mass ratio and so once again there
is an effective Lagrangian to order $v^4$.  It seems likely that a
similar phenomenon may occur here.

Another issue which we have not fully addressed is the
polarization dependence of scattering in the moduli space
approximation. The string amplitude \seventeen\ indicates that
there are new contributions to the scattering when the polarization vectors
lie partially in the scattering plane. In string theory this dependence
arises due to the presence of fermion zero modes in the
vertex operators for scattering of string states, and it seems likely
that it can be understood in the moduli space approximation
by studying scattering on the supermoduli space
of heterotic string configurations.

\bigskip\centerline{\bf Acknowledgements}\nobreak
We would like to thank C.  Callan, S. Ferrara, E. Martinec, T. Samols, and A.
Strominger for
discussions. This work was supported in part by
NSF
Grant No.~PHY90-00386.  J.P.G.\ is supported by a grant from the
Mathematical Discipline Center of the Department of Mathematics,
University of Chicago. J.H.\ also acknowledges the support of NSF PYI
Grant No.~PHY-9196117.

\listrefs
\end